# Ternary Bismuthide SrPtBi$_2$: Computation and Experiment in Synergism to Explore Solid-State Materials

Xin Gui,[a] Xin Zhao,[b*] Zuzanna Sobczak,[c] Cai-Zhuang Wang,[b] Tomasz Klimczuk,[c] Kai-Ming Ho,[b] Weiwei Xie [a*]

[a] Department of Chemistry, Louisiana State University, Baton Rouge, LA, USA 70803

[b] Department of Physics and Astronomy, Iowa State University and Ames Laboratory, US Department of Energy, Ames, IA, USA 50011

[c] Faculty of Applied Physics and Mathematics, Gdansk University of Technology, Narutowicza 11/12, Gdansk, Poland 80–233

**ABSTRACT:** A combination of theoretical calculation and the experimental synthesis to explore the new ternary compound is demonstrated in the Sr-Pt-Bi system. Since Pt-Bi is considered as a new critical charge-transfer pair for superconductivity, it inspired us to investigate the Sr-Pt-Bi system. With a thorough calculation of all the known stable/metastable compounds in the Sr-Pt-Bi system and crystal structure predictions, the thermodynamic stability of hypothetical stoichiometry, SrPtBi$_2$, is determined. Followed by the high temperature synthesis and crystallographic analysis, the first ternary bismuthide in Sr-Pt-Bi, SrPtBi$_2$ was prepared and the stoichiometry was confirmed experimentally. SrPtBi$_2$ crystallizes in the space group *Pnma* (S.G. 62, Pearson Symbol *oP*48), which matches well with theoretical prediction using an adaptive genetic algorithm (AGA). Using first-principles calculations, we demonstrate that the orthorhombic structure has lower formation energies than other 112 structure types, such as tetragonal BaMnBi$_2$ (CuSmP$_2$) and LaAuBi$_2$ (CuHfSi$_2$) structure types. The bonding analysis indicates the Pt-Bi interactions play a critical role in structural stability. The physical properties measurements show the metallic properties with low electron-phonon interactions at the low temperature, which agrees with the electronic structure assessment.

## INTRODUCTION

Exploring novel functional solid-state materials, which are key to technology innovation, is a long-standing goal in materials chemistry and physics. Because of various possibilities of combining atoms, discovering novel materials has long been found to be a time-consuming process. Traditionally, prediction of new compounds heavily relies on empirical or semi-empirical rules, such as electron counts and packing rules.[1-3] Current strategies involving informatics, machine learning, and using of materials databases, such as Materials Project and AFLOWLIB, opened up an efficient approach to expedite the search for functional solid state materials.[4-7] Briefly, truly representative atomic combinations will be selected based on the highly performed computation. These atomic combinations will be evaluated by experimental synthesis to see which one is more stable and reliable. The process with a combination of experiments and theory has effectively reduced the scope for finding new materials.[8]

As non-toxic heavy metals, bismuth shows the strong spin-orbital coupling effects, which provide bismuthides many unique and desirable properties for optoelectronic, thermoelectric and electronic device applications.[9] For example, three-dimensional Dirac fermions have been detected in the topological Dirac semimetal, Na$_3$Bi.[10,11] Such materials can be used as transistors and other electronic devices.[12] Usually, simple Bi-containing semiconductors or semimetals can be predicted using chemical valence arguments or Zintl-Klemm concepts.[13-15] Half-Heusler RPtBi (R= rare earth elements) compounds following 18e- rules show intriguing properties such as superconductivity in LuPtBi.[16-18] The Zintl-Klemm concepts can be employed for interpreting the charge balance in BaMnBi$_2$, which is an antiferromagnetic Dirac material coexisting with superconductivity at high pressure.[19] BaMnBi$_2$ can be formulated as

$Ba^{2+}Mn^{2+}Bi^-Bi^{3-}$ with Bi-Bi square planar in the structure.[20] However, if a system turns more complex with small difference in electronegativity among elements, chemical valence rules may fail in predicting crystal structures and properties. It is even harder to rationalize some subtle structural distortions or atomic distributions due to the complexity. For example, ternary rare-earth gallium bismuthide, $LaGaBi_2$ off the chemical balance, crystallizes in the hexagonal structure with 24 atoms per unit cell and $La_6$ trigonal prisms centered by Bi atoms.[21]

Recently, we discovered the superconductivity in new monoclinic $BaPt_2Bi_2$. The research indicates the Pt-Bi antibonding is significant for inducing superconductivity. Thus, we extended our study to the Sr-Pt-Bi system, which was unexplored before. To accelerate the targeted synthesis, we employed the crystal structure prediction and total energy calculation firstly to determine the thermodynamic stable stoichiometry. Using the predicted stoichiometry, we successfully synthesized the new $SrPtBi_2$. Unlike other tetragonal 112-$CuSmP_2$ ($BaMnBi_2$)[19] or $CuHfSi_2$ ($LaAuBi_2$)[22], $SrPtBi_2$ adpots the orthorhombic structure with space group *Pnma*, which is analogue to $YNiSn_2$.[23] Electronic structure calculations based on the experimental crystal information have confirmed the structural preference and chemical stability of orthorhombic $SrPtBi_2$. The heat capacity measurements show the metallic properties of $SrPtBi_2$ with weak electron-phonon interactions at the low temperature.

**EXPERIMENTAL SECTION**

**Synthesis** Polycrystalline samples of $SrPtBi_2$ were obtained *via* high-temperature solid-state method. Elemental strontium (> 99%, rod, Alfa Aesar), platinum powder (99.98%, ~ 60 meshes, Alfa Aesar) and ground bismuth powder (99.999%, lump, Alfa Aesar) were pelletized in the glovebox with a ratio of 1:1:2 and placed into an alumina crucible which was subsequently sealed into an evacuated ($10^{-5}$ torr) quartz tube. The sample was heated to 900 °C at a rate of 1 °C/min and annealing at 900 °C for 2 days. After that, the tube was quenched in the air. It yielded a mixture of $SrPtBi_2$ and minor $PtBi_2$ impurity. All products are sensitive in the moisture.

**Phase Analyses and Structure Determination** All $SrPtBi_2$ samples were ground finely and examined by a Rigaku MiniFlex 600 powder X-ray diffractometer equipped with Cu $K_\alpha$ radiation ($\lambda$=1.5406 Å, Ge monochromator). A Bragg angle ranging from 5° to 90° in a step of 0.010° at the rate of 0.08°/min was taken to get a precise pattern. The lattice parameters and phase composition refinements were accomplished using JANA 2006 with LeBail model.[24,25] Multiple crystals (~20 pieces) with roughly 0.02 mm diameter were picked up to carry out single crystal X-ray diffraction in case of heterogeneity. A Bruker Apex II diffractometer equipped with Mo radiation ($\lambda_{K\alpha}$= 0.71073 Å) was utilized to determine the structure at room temperature. The sample protected with glycerol was mounted on a Kapton loop. In order to guarantee the accuracy, seven different angles were chosen to take the measurement at room temperature with an exposure time of 15 seconds per frame while the width of scanning was 0.5°. The direct methods and full-matrix least-squares on $F^2$ models along with SHELXTL package were employed to solve the crystal structure.[26] Data acquisition was made via Bruker SMART software which made corrections for Lorentz and polarization effects as well.[27] On the basis of face-index modeling, numerical absorption corrections were approached by *XPREP*.[28]

**Scanning Electron Microscopy (SEM)** Characterization was performed using a high-vacuum scanning electron microscope (JSM-6610 LV) and Energy-Dispersive Spectroscopy (EDS). Samples were quickly moved from the glovebox and mounted on the carbon tape because of the air-sensitivity before loading into the SEM chamber. Multiple points and areas were examined in each phase within multiple grains of a specimen. The samples were examined at 15 kV. Spectra were collected for 100 seconds.

**Heat Capacity Measurements** The specific heat measurements were performed using a physical properties measurement system (PPMS) on pieces of SrPtBi$_2$ polycrystalline samples in Figure S1. The measurements were operated over a temperature range of 1.8-300 K without applied magnetic field.

## COMPUTATIONAL METHODS

**Adaptive Genetic Algorithm (AGA)** The crystal structure search was performed using the AGA method.[29, 30] AGA integrates auxiliary interatomic potentials and first-principles calculations together in an adaptive manner to ensure the high efficiency and accuracy. Interatomic potentials based on the embedded-atom method (EAM) were selected as the auxiliary classical potentials in the study of Sr-Pt-Bi system.[31] Structure searches of SrPtBi$_2$ were performed for unit cell sizes of 2, 3, 4, 6, and 12 formula units. During the searches, no symmetry constraint was applied and the total structure population was set to be 128. Convergence was considered to be reached when the lowest energy in the structure pool remained unchanged for 500 consecutive generations. At the end of each classical GA search, the lowest-energy structures were selected to perform first-principles calculations according to the AGA procedure, whose energies, forces, and stress were used to adjust the parameters of the EAM potential using the *Potfit* code.[32, 33] Finally, all of the structures used to tune the interatomic potentials during the AGA search were collected for higher-accuracy optimization by first-principles calculations.

**Vienna Ab initio Simulation Package (VASP)** First-principles calculations were carried out using density functional theory (DFT) within a generalized gradient approximation (GGA) by VASP code.[34, 35] Projector augmented-wave method[36] was used to describe the valence configuration: $4s^24p^65s^2$ for Sr, $5d^96s^1$ for Pt, and $5d^{10}6s^26p^3$ for Bi. The GGA exchange-correlation energy functional parametrized by Perdew, Burke, and Ernzerhof was used.[37] Plane-wave basis was used with a kinetic energy cutoff of 520 eV. Monkhorst−Pack's scheme was adopted for Brillouin zone sampling with a *k*-point grid of $2\pi \times 0.05$ Å$^{-1}$ during the AGA searches.[38] In the final structure refinements of the collected SrPtBi$_2$ structures from the AGA searches as well as the known binary/ternary compounds in the Sr-Pt-Bi system, a denser grid of $2\pi \times 0.025$ Å$^{-1}$ was used, and the ionic relaxations stopped when the forces on every atom became smaller than 0.01 eV/Å.[39] To characterize the thermodynamical stability, the formation energy of any given structure Sr$_m$Pt$_n$Bi$_p$ was calculated using face-centered cubic (FCC) Pt, FCC Sr and rhombohedral Bi as references, i.e. $E_F(Sr_mPt_nBi_p) = [E(Sr_mPt_nBi_p) - m*E(Sr) - n*E(Pt) - p*E(Bi)]/(m+n+p)$.

**Tight-Binding, Linear Muffin-Tin Orbital-Atomic Spheres Approximation (TB-LMTO-ASA)** Calculations of the electronic and bonding features were performed by TB-LMTO-ASA using the Stuttgart code.[40-41] Exchange and correlation were treated by the local density approximation (LDA).[42] In the atomic sphere approximation method, the space is filled with overlapping Wigner-Seitz (WS) spheres.[43] During the calculation of SrPtBi$_2$, the empty spheres are used to make the overlap of WS spheres limited to no larger than 16%. The WS radii are: 2.14-2.17 Å for Sr; 1.50-1.53 Å for Pt; and 1.63-1.79 Å for Bi. The basis set for the calculations included Sr 5*s*, 4*d*, Pt 6*s*, 6*p*, 5*d*, and Bi 6*s*, 6*p* wavefunctions. The convergence criterion was set to 0.1 meV. A mesh of 60 *k* points in the irreducible wedge of the first Brillouin zone was used to obtain all integrated values, including the density of states (DOS), band structure, and Crystal Orbital Hamiltonian Population (COHP) curves.[44]

## RESULTS AND DISCUSSION

**Construction of the Convex Hull** Convex hull of the Sr-Pt-Bi system was constructed to investigate the stability of different stoichiometry, as plotted in Figure 1. Because there is currently no ternary phase existing in the Sr-Pt-Bi system, its convex hull is determined solely by elemental Sr, Pt, Bi and the binary compounds. Thus, we performed a thorough calculation on all the known binary phases among Sr, Pt, and Bi. During our calculation, we also discovered several metastable binary phases in the Sr-Bi system, such as $Sr_3Bi_2$, $Sr_2Bi$, and $Sr_3Bi$ marked in blue circle. The thermodynamic stability of $SrPtBi_2$ is determined by the red Gibbs triangle in Figure 1, i.e.

$$SrPtBi_2 \rightarrow x * PtBi_2 + y * Sr_2Bi_3 + z * SrPt_2$$

When the energy of the left hand side in the above reaction (energy of $SrPtBi_2$) is lower than that of the right hand side (sum of the energies of $x*PtBi_2$, $y*Sr_2Bi_3$, $z*SrPt_2$), $SrPtBi_2$ is considered to be thermodynamically stable. Our calculation shows that the formation energy of the right hand side, i.e. the energy on the convex hull at the composition of $SrPtBi_2$ is -574.74 meV/atom.

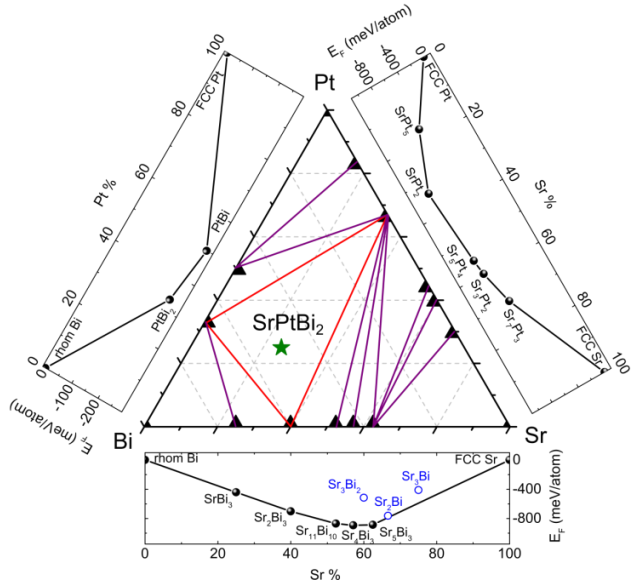

**Figure 1.** Convex hull of the formation energies in the Sr-Pt-Bi system. Black balls (and black triangles in the ternary phase diagram) represent stable compounds, while blue points indicate the metastable phases. The purple lines in the ternary phase diagram are projections of the convex hull construction into compositional space, which forms Gibbs triangles.

**Exploration of the $SrPtBi_2$ Structural Space** Our crystal structure searches of $SrPtBi_2$ were performed using unit cell sizes of 2, 3, 4, 6, and 12 formula units. The calculated formation energies of the obtained structures are plotted in Figure 2(a) together with the hypothetical structures ($BaMnBi_2$ and $LaAuBi_2$) and the energy on the convex hull. The structures from the search are classified in terms of the Pt-Bi coordination number. It can be seen that structures below the convex hull are found for $SrPtBi_2$. On the other hand, the hypothetical models with $BaMnBi_2$ and $LaAuBi_2$-type structures have energies of ~35 meV/atom higher than the convex hull. The lowest-energy $SrPtBi_2$ structure is found be in the space group of *Pnma* with Pt@$Bi_5$ square pyramidal Bi5 coordination. It has ~50meV/atom lower formation energy than the one with Pt@$Bi_4$ tetrahedral Bi4 coordination in $LaAuBi_2$-type and $BaMnBi_2$-type.

Phonon spectrum of the orthorhombic $SrPtBi_2$ was calculated to investigate its dynamical stability. Calculations were performed using a supercell approach provided by the Phonopy code [45] where supercells with sizes of 144 atoms (1×3×1) and *k*-mesh of 2×2×2 were used. The phonon density-of-states (DOS) results are plotted in Figure 4(b). It can be seen that there are no imaginary phonon frequencies in this structure, indicating that it is dynamically stable. Thus far, we demonstrate that $SrPtBi_2$ is a promising stoichiometry to discover new thermodynamically and dynamically stable compound.

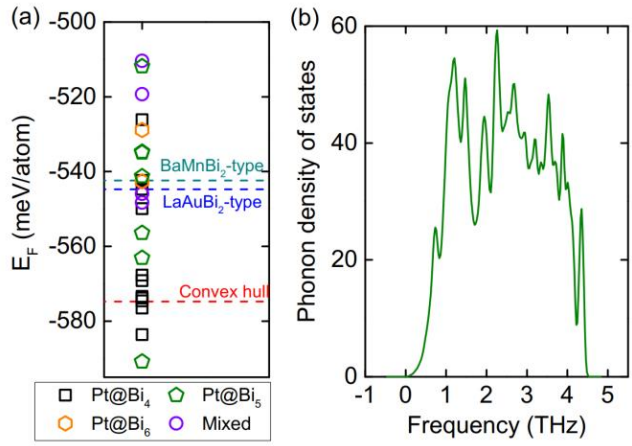

**Figure 2.** (a) Formation energies of the SrPtBi$_2$ crystal structures obtained from the AGA searches. The structures are classified by their Pt-Bi coordination, where "Mixed" represents structures with more than one type of Pt-Bi polyhedra. Energies of the hypothetical structures, i.e. the BaMnBi$_2$- and LaAuBi$_2$-types (cyan and blue dash lines) and the energy on the convex hull at the composition of SrPtBi$_2$ are plotted as references (red dash line). (b) Phonon density of states of the lowest-energy structure.

**Experimental Confirmation of SrPtBi$_2$** The first ternary compound in Sr-Pt-Bi system, SrPtBi$_2$, was produced by high temperature synthesis. To obtain the detailed chemical stoichiometry and atomic distributions, the single crystal X-ray diffraction was used to determine the crystal structure. SrPtBi$_2$ crystallizes in orthorhombic LuNiSn$_2$-type structure with the space group *Pnma*,[23] same as our crystal structure prediction. The resulting structural analysis including the site occupancies, isotropic thermal displacements and atomic positions was shown in Table 1 and Table 2. The flexible and mixed site occupancy models were utilized to test and confirm the atomic order and stoichiometry in SrPtBi$_2$. Because of the heavy Pt and Bi elements, the site splitting was considered for the refinement, but no site splitting was observed. The three crystallographic independent Pt atoms form Pt@Bi$_5$ square pyramidal polyhedral, which share the edges and vertex with neighboring polyhedral. The Sr atoms were in the voids of Pt-Bi frames according to Figure 3(*a*). As shown in Figure 3(*b*) and by the formation energies in 2(*a*), among the potential structures with 112 stoichiometry adopted by SrPtBi$_2$, Pt@Bi$_5$ polyhedra in space group *Pnma* is determined to hold ~ 7 and ~ 48 meV per atom lower energy than Pt@Bi$_4$ and Pt@Bi$_6$ polyhedra, respectively.

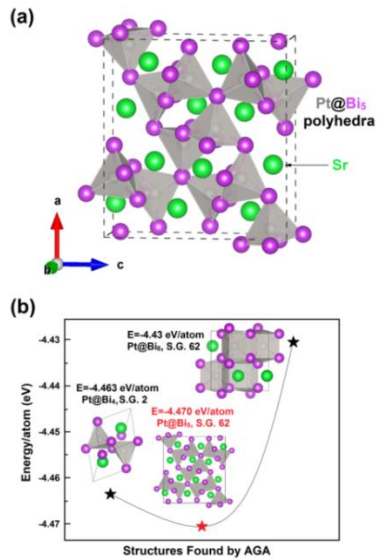

**Figure 3.** (*a*) Crystal structure of SrPtBi$_2$ emphasizing the Pt@Bi$_5$ square pyramidal Bi coordinations. (*b*) Schematic figure showing the total energies of SrPtBi$_2$ with different structures screened by AGA.

**Table 1.** Single crystal crystallographic data for SrPtBi$_2$ at 296 (2) K

| Refined Formula | SrPtBi$_2$ |
|---|---|
| F.W. (g/mol) | 700.67 |
| Space group; Z | *Pnma*; 12 |
| $a$(Å) | 17.072 (4) |
| $b$(Å) | 4.893 (1) |
| $c$(Å) | 15.806 (4) |
| V (Å$^3$) | 1320.4 (6) |
| Extinction Coefficient | 0.00008 (1) |
| θ range (deg) | 1.756-33.169 |
| No. reflections; $R_{int}$ | 24272; 0.1035 |
| No. independent reflections | 2770 |
| No. parameters | 74 |
| $R_1$: $\omega R_2$ ($I>2\sigma(I)$) | 0.0511; 0.0876 |
| $R_1$: $\omega R_2$ (all $I$) | 0.0687; 0.1071 |
| Goodness of fit | 0.961 |
| Diffraction peak and hole (e$^-$/ Å$^3$) | 5.552; -3.756 |

**Table 2.** Atomic coordinates and equivalent isotropic displacement parameters of SrPtBi$_2$ system. ($U_{eq}$ is defined as one-third of the trace of the orthogonalized $U_{ij}$ tensor (Å$^2$))

| Atom | Wyck. | Occ. | x | y | z | $U_{eq}$ |
|---|---|---|---|---|---|---|
| Bi1 | 4*c* | 1 | 0.0335(1) | ¼ | 0.2610(1) | 0.0182(2) |
| Bi2 | 4*c* | 1 | 0.0429(1) | ¼ | 0.5442(1) | 0.0259(3) |
| Bi3 | 4*c* | 1 | 0.2882(1) | ¼ | 0.3907(1) | 0.0122(2) |
| Bi4 | 4*c* | 1 | 0.3136(1) | ¼ | 0.6726(1) | 0.0118(2) |
| Bi5 | 4*c* | 1 | 0.3332(1) | ¼ | 0.1234(1) | 0.0131(2) |
| Bi6 | 4*c* | 1 | 0.4773(2) | ¼ | 0.5776(2) | 0.0114(2) |
| Sr1 | 4*c* | 1 | 0.1200(1) | ¼ | 0.7661(1) | 0.0171(6) |
| Sr2 | 4*c* | 1 | 0.1524(1) | ¼ | 0.0273(1) | 0.0114(5) |
| Sr3 | 4*c* | 1 | 0.3559(1) | ¼ | 0.8896(2) | 0.0164(6) |
| Pt1 | 4*c* | 1 | 0.2021(1) | ¼ | 0.2425(1) | 0.0157(2) |
| Pt2 | 4*c* | 1 | 0.2036(1) | ¼ | 0.5426(1) | 0.0154(2) |
| Pt3 | 4*c* | 1 | 0.4533(2) | ¼ | 0.3990(2) | 0.0133(2) |

    The refined powder X-ray diffraction pattern is shown in Figure 4. The LeBail fitting was used to refine all lattice parameters. PtBi$_2$ was found as impurity (~10%) existing in the sample. Without refinement of the atomic sites or displacement parameters of all atoms, the obtained profile residuals $R_p$ is 5.98% with weighted profile residuals $R_{wp}$ 9.52%. Refined lattice parameters for SrPtBi$_2$ (orthorhombic symmetry, $a$ = 17.093(2) Å; $b$ = 4.8739(4) Å; $c$ = 15.725(1) Å) were found consistent with single crystal X-ray diffraction data. In comparison, our theoretical calculation using DFT-GGA slightly overestimate the lattice parameters and gives $a$ = 17.334 Å; $b$ = 4.949 Å; $c$ = 16.012 Å at equilibrium conditions. The chemical composition was further confirmed as Sr$_{1.1(1)}$Pt$_{1.0(1)}$Bi$_{1.8(2)}$ by using Scanning Electron Microscopy (SEM), as shown in Figure S2.

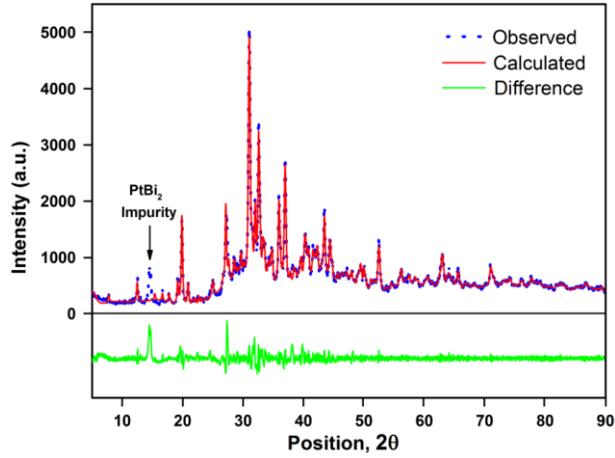

**Figure 4.** Powder XRD pattern for SrPtBi$_2$. The red dots, blue line and green line represent calculated Powder XRD pattern, observed Powder XRD pattern and the intensity difference between calculated and observed pattern, respectively. PtBi$_2$ peaks were found as minor impurity (~10%).

**Electronic Structure and Bonding Features of SrPtBi$_2$** In order to understand the structural stability from the bonding aspect, electronic structures were calculated and shown in Figure 5. We emphasized the range around Fermi level between the energy of -4.0 eV and 2.0 eV. In the DOS, we can easily see the 5*d* orbitals from Pt atoms contribute most below -3.0 eV, which means the electrons on 5d orbitals are relatively localized. Above -3.0eV, the states are dominated by the hybridization of electrons on Bi-*s* and *p* orbitals, Pt-*s* and *d* orbitals, and Sr-*s* orbital, in particular around the Fermi level. To specify the atomic interactions around Fermi level, the Crystal Orbital Hamilton Populations (-COHP) calculations were generated in Figure 5(Middle), in which the positive part (+) indicates the bonding interaction while the negative part (-) indicates anti-bonding interactions. The outcome of the –COHP calculation demonstrates that the Pt-Bi interactions play a significant role in the structural stability and Fermi level locates on the non-bonding region in Pt-Bi interactions. Around Fermi level, the interplay between Bi-Bi anti-bonding and Sr-Bi bonding interaction governs the stability of SrPtBi$_2$. Even the peak around Fermi level is observed in DOS, the band structure calculation shows that the bands are localized near the Fermi level with no saddle points which hints that superconductivity can rarely been found in this compound.

Similar with the integrating electronic DOS which gives the number of electrons in the system, the integrated COHP hints towards the bond strength.[46-47] To compare the atomic interactions within each model and understand the bonding strengths in different 112 structure models, the corresponding integrated COHP values are calculated and listed in Table 3. Accordingly, in experimental SrPtBi$_2$ structure, the Pt-Bi interactions contribute most (~69%) in the structural stability while there exists no Pt-Pt interactions. On the other hand, the Pt-Bi interactions in LaAuBi$_2$ and BaMnBi$_2$ models are dramatically decreasing. The decreasing of Pt-Bi bonding strength leads to the instability of SrPtBi$_2$ in other two hypothetical models. Moreover, the changing trend for total energies is consistent with the Pt-Bi bonding strength of SrPtBi$_2$ in different models. In this regards, the Pt-Bi interactions play a significant role in determining the stability of the compound. It is worth noticing that the Pt-Pt interactions also contribute critically in the superconducting BaPt$_2$Bi$_2$ but they are not observed in SrPtBi$_2$. It may be the reason of SrPtBi$_2$ not being superconductor since the superconductivity requires the interplay between Pt-Bi and Pt-Pt interactions.

**Table 3.** Integrated COHP per formula for SrPtBi$_2$ in experimental, LaAuBi$_2$-, and BaMnBi$_2$-type structures.

|       | SrPtBi$_2$-type | | LaAuBi$_2$-type | | BaMnBi$_2$-type | |
|-------|-------|---------|-------|---------|-------|---------|
|       | ICOHP | ICOHP(%) | ICOHP | ICOHP(%) | ICOHP | ICOHP(%) |
| Pt-Bi | 2.24 | 68.70 | 3.31 | 40.66 | 1.38 | 28.30 |
| Bi-Bi | 0.25 | 7.53 | 1.66 | 20.39 | 1.47 | 30.14 |
| Sr-Bi | 0.62 | 18.90 | 2.09 | 25.68 | 1.09 | 22.38 |
| Sr-Pt | 0.16 | 4.87 | 1.08 | 13.27 | 0 | 0 |
| Pt-Pt | 0 | 0 | 0 | 0 | 0.94 | 19.17 |
| Total | 3.27 | 100 | 8.14 | 100 | 4.88 | 100 |

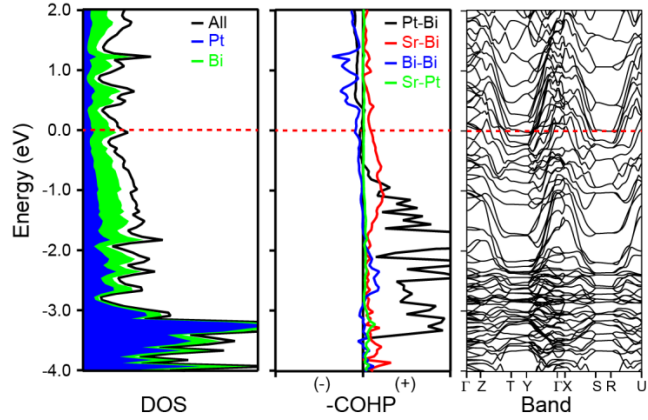

**Figure 5.** Electronic structure calculation of SrPtBi$_2$ from the energy of -4.0 eV to 2.0 eV. (Left) Density of States; (Middle) Crystal Orbital Hamilton Populations (–COHP) calculations (+ indicates the bonding interactions and – indicates the antibonding interactions); (Right) Band structure calculations.

**Heat Capacity Measurements** Temperature-dependent specific heat measurements were carried out as presented in Figure 6(main panel), which plots $C_p$ vs T in zero applied field from 1.8 to 300 K. The inset of Figure 6 shows a plot of $C_p/T$ vs $T^2$ which was plotted to equation, $\frac{C_p}{T} = \gamma + \beta T^2$, where $\beta T^3$ is the phonon contribution and $\gamma T$ is the electronic contribution to the specific heat. The Sommerfeld parameter, $\gamma$, was calculated to be 6.4(3) mJ/mol-f.u.K$^2$. The small gamma indicates the metallic properties with strong electron-electron coupling and weak electron-phonon interactions. The Debye temperature $\Theta_D$ can then be calculated with the following equation, $\Theta_D = (\frac{12\pi^4}{5\beta} nR)^{1/3}$, where R is the gas constant and $n=4$ for SrPtBi$_2$.[48] Based on this Debye model, the Debye temperature is calculated to be 173 K, which also indicates the weak electron-phonon interactions. No anomalies were observed, such as phase transition or superconductivity.

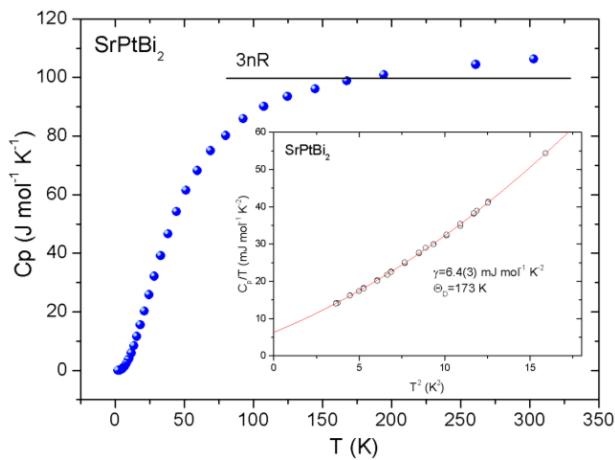

**Figure 6.** Heat capacity measurements of SrPtBi$_2$. (Main Panel) Molar specific heat capacity vs temperature without applied magnetic field. (Insert)

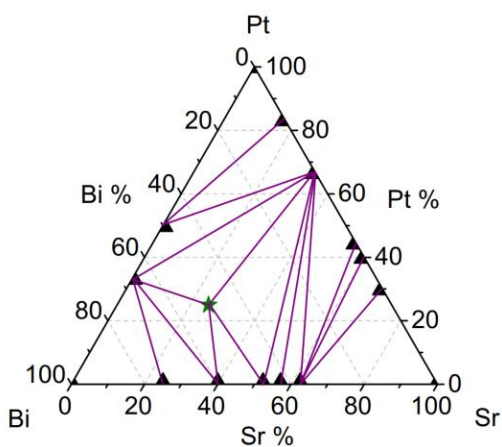

**Figure 7.** Updated convex hull of the Sr-Pt-Bi system after including the new SrPtBi$_2$ phase discovered in the current work.

## CONCLUSIOINS

The first ternary phase in Sr-Pt-Bi system, SrPtBi$_2$, has been designed and obtained by coupling theoretical computation and experiments. SrPtBi$_2$ phase was synthesized and structurally characterized. It exhibits in the orthorhombic structure with space group *Pnma*. First principles electronic and phonon structure calculations substantiate the chemical stability of the new compound and indicate that no superconductivity may be observed in the phase. The heat capacity measurements confirmed the metallic properties. No superconductivity was detected down to 1.8K. With our newly discovered SrPtBi$_2$ phase, the phase diagram of the Sr-Pt-Bi system can be correspondingly updated as shown in Figure 7. It can be seen that the existence of SrPtBi$_2$ (green star) alters the equilibrium stable phases at the compositions which are not the stable nodes, which will aid the search for the new superconductors in Sr-Pt-Bi system with the interplay between Pt-Pt and Pt-Bi interactions.

## ASSOCIATED CONTENT

**Supporting Information**

This material is available free of charge via the ACS Publications website at http://pubs.acs.org.

SEM images of SrPtBi$_2$ and heat capacity measurement

## AUTHOR INFORMATION

**Corresponding Author**

Weiwei Xie (weiweix@lsu.edu)

Xin Zhao (xzhao@iastate.edu)

**Notes**

The authors declare no competing financial interest.

**ACKNOWLEDGMENT**

X.G. and W.X. deeply thank the support from Louisiana State University and the Louisiana Board of Regents Research Competitiveness Subprogram (RCS) under Contract Number LEQSF(2017-20)-RD-A-08 and the Shared Instrument Facility (SIF) at Louisiana State University for the SEM-EDS. Work at Ames Laboratory was supported by the U.S. Department of Energy (DOE), Office of Science, Basic Energy Sciences, Materials Science and Engineering Division including a grant of computer time at the National Energy Research Scientific Computing Centre in Berkeley, CA. Ames Laboratory is operated for the U.S. DOE by Iowa State University under contract # DE-AC02-07CH11358.

**Table of Content**

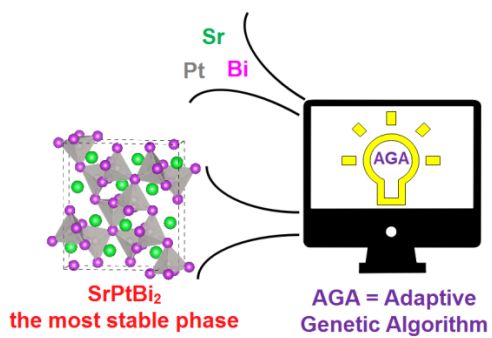

A combination of theoretical calculation and the experimental synthesis to explore the new ternary compound is demonstrated in the Sr-Pt-Bi system. With a thorough calculation of all the known stable/metastable compounds in the Sr-Pt-Bi system and crystal structure predictions, the thermodynamic stability of hypothetical stoichiometry, SrPtBi$_2$, is determined. Followed by the high temperature synthesis and crystallographic analysis, the first ternary bismuthide in Sr-Pt-Bi, SrPtBi$_2$ was prepared and the stoichiometry was confirmed experimentally. SrPtBi$_2$ crystallizes in the space group *Pnma* (S.G. 62, Pearson Symbol *oP*48), which matches well with theoretical prediction using an adaptive genetic algorithm (AGA).